\documentclass[final,5p,times,twocolumn]{elsarticle}
\usepackage{graphicx}
\usepackage{amssymb}
\usepackage{amsmath}
\journal{Polyhedron}

\begin{document}

\begin{frontmatter}
\title{Quantum Fluctuations and Long-Range Order in Molecular Magnets}

\author[1]{P. Subedi\corref{cor1}}
\ead{pradeep.subedi@nyu.edu}
\author[2]{Bo Wen}
\author[3]{Y. Yeshurun}
\author[2]{M. P. Sarachik}
\author[4]{A. J. Millis}
\author[1]{A. D. Kent\corref{cor1}}
\ead{andy.kent@nyu.edu}
\address[1]{Department of Physics, New York University, New York, New York 10003, USA}
\address[2]{Department of Physics, City College of NY, CUNY, New York, New York 10031, USA}
\address[3]{Department of Physics, Bar-Ilan University, Ramat-Gan 52900, Israel}
\address[4]{Department of Physics, Columbia University, New York, New York 10027, USA}
\cortext[cor1]{Corresponding authors}

\begin{abstract}
We review our studies of the effect of transverse fields on the susceptibility and magnetization of single crystals of the prototype single molecule magnet (SMM), Mn$_{12}$-acetate, and of a new high-symmetry variant, Mn$_{12}$-acetate-MeOH. SMM single crystals can exhibit long range ferromagnetic order associated with intermolecular dipole interactions. Transverse fields increase quantum spin fluctuation and quantum tunneling of the magnetization suppressing long range order. However, we have found that suppression of the Curie  temperature by a  transverse field in Mn$_{12}$-acetate is far more rapid than predicted by the Transverse-Field Ising Ferromagnetic Model (TFIFM). It appears that solvent disorder in Mn$_{12}$-acetate that results in an intrinsic distribution of small discrete tilts of the molecular magnetic easy axis from the global easy axis of the crystal ($\approx \pm 1^\circ$) gives rise to a distribution of random-fields that further suppresses long-range order. Semiquantitative agreement with the predictions of a Random-Field Ising Ferromagnet Model is found. Subsequent susceptibility studies we have conducted of the high symmetry Mn$_{12}$ variant, Mn$_{12}$-acetate-MeOH, with the same spin structure and similar lattice constants but without the same solvent disorder as Mn$_{12}$-acetate, agrees with the TFIFM. An important implication of our studies is that long-range order in these two chemically very similar SMMs are described by distinct physical models.
\end{abstract}
\end{frontmatter}

\section{Introduction}
\label{Intro}
The single molecule magnets (SMMs) are metalorganic compounds composed of transition-metal ions and organic ligand molecules. At low temperature, each molecule has a large ground-state spin and a strong uniaxial anisotropy, leading to Ising-like behavior, with its spin oriented either up or down along the easy axis. The molecules form van der Waals crystals and the intermolecular exchange interactions between molecules are typically quite small. This has led to the commonly used approximation that the large spins on each molecule are independent and do not interact with one another. Indeed, much of early research on SMMs did not consider intermolecular interactions or collective phenomena. Such studies include the famous observation of quantum tunneling of magnetization in Mn$_{12}$-acetate \cite{JonathanPRL1996}, and the study of Berry phase oscillations in Fe$_8$ \cite{Wernsdorfer1999}. However, the molecules do interact with each other and the dominant interaction is often magnetic dipolar interaction and, while the strength of the intermolecular dipole interactions is much less than the scale of the anisotropy energy of the molecules, there are quite notable effects of intermolecular interactions. This has lead to a renewed interest in studies of collective phenomena due to dipolar interactions in SMM single crystals \cite{LuisPRL2005,Subedi2010,Burzuri,Subedi2012}.

For spins on an ordered lattice, dipolar interactions lead to long range order with a ferromagnetic or antiferromagnetic ground state that depends on the lattice structure and sample shape. In several SMMs, finite temperature transitions to a dipolar ferromagnetic state have been observed or predicted theoretically \cite{LuisPRL2005,PhysRevLett.90.017206,PhysRevLett.93.117202,Millis09,evangelisti:167202,FernandezPRB2000,garanin:174425}. In Mn$_{12}$-acetate, a transition to dipolar ferromagnetism was inferred from neutron scattering experiments by Luis et al. \cite{LuisPRL2005}. This finding was supported by Monte Carlo simulations \cite{FernandezPRB2000}. Calculations based on the Mean Field Approximation (MFA) by Garanin and Chudnovsky \cite{garanin:174425} and Millis et al. \cite{Millis09} have predicted the existence of an ordered state in elongated crystals of Mn$_{12}$-acetate at low temperature.  Our measurements of the longitudinal magnetization and susceptibility of Mn$_{12}$-acetate are consistent with the occurrence of a transition to dipolar ferromagnetism at finite temperature \cite{Subedi2010}. In our experiments we apply transverse fields to induce quantum fluctuations (i.e. increase the rate of quantum tunneling of magnetization) and study the interplay between intermolecular interactions and quantum fluctuations by observing the field dependence of the magnetic susceptibility \cite{Subedi2010,Subedi2012}. 

In this article we review our experimental and theoretical studies of interactions and long-range order in molecular magnets. In Section 2, we introduce the molecular magnets that we study. In Section 3 we discuss our experimental setup and results. In the following section we present our theoretical model. In section 5 we present a comparison between theoretical calculations and experimental data. And, finally, we present our conclusions and perspectives. 

\section{Single Molecule Magnets}
\label{SMM}
\subsection{\label{ssec:level1}Mn$_{12}$-acetate}
One of most studied SMMs is Mn$_{12}$-acetate, [Mn$_{12}$O$_{12}$(O$_{2}$CCH$_{3}$)$_{16}$(H$_{2}$O)$_{4}$]$ \cdot $2CH$_{3}$CO$_{2}$H$\cdot $ 4H$_{2}$O. Each Mn$_{12}$ molecule behaves as a nanomagnet with spin $S = 10$ oriented along the crystallographic \textit{c} axis by a strong uniaxial magnetic anisotropy $ DS^2=65$\,K, where $D$ is the first order uniaxial anisotropy constant. Mn$_{12}$-acetate crystallizes in a body centered tetragonal lattice (space group $I\bar{4}$) with unit cell parameters at $83$ K of $a = b = 17.1668(3)$ \AA, $c = 12.2545(3)$ \AA, molecules per unit cell(Z) = 2, $V = 3611.39$ \AA$^3$ \cite{corniaACC2002}. The magnetic centers are well separated by the organic ligands and intermolecular exchange is small compared to the intermolecular magnetic dipole interactions \cite{Park02}, $E_\mathrm{dip}=(g\mu_{B})^{2}S^{2}/(a^{2}c)\sim 0.08$ K, which is one order of magnitude larger than intramolecular hyperfine interactions. In the absence of applied fields, magnetic ordering with a Curie temperature of about $0.9$ K \cite{garanin:174425,Millis09} has been predicted.

 \subsection{\label{ssec:level2}Mn$_{12}$-acetate-MeOH}
Mn$_{12}$-ac-MeOH, [Mn$_{12}$O$_{12}$(O$_2$CMe)$_{16}$(MeOH)$_4$]$\cdot$MeOH, is a new high-symmetry variant of the original SMM Mn$_{12}$-ac. Each molecule in Mn$_{12}$-ac-MeOH has the same ground state spin of S = 10 and similar anisotropy of $ DS^2=66.7$ K as the molecules in original Mn$_{12}$-ac. Mn$_{12}$-ac-MeOH also crystallizes in the space group $I\bar{4}$ with unit cell parameters at 173 K of $a = b = 17.3500(18)$ \AA, $c = 11.9971(17)$ \AA, Z = 2, $V = 3611.4$ \AA$^3$. \cite{Stamatatos,PhysRevB.80.094408, Bian2005}. All the parameters are nearly identical to Mn$_{12}$-acetate.    
  
Although the two compounds have similar lattice parameters, spin and anisotropy, there are some crucial differences relevant to this study, which originate from the organic ligand and solvent molecules that form the local molecular environment. In Mn$_{12}$-ac-MeOH, the four terminal water molecules of Mn$_{12}$-ac are replaced by terminal methanol molecules, the two acetic acid and four water solvent molecules are replaced by only one methanol that resides on a symmetry element making the overall structure highly symmetric and leading the crystal to retain the molecular $S_{4}$  symmetry \cite{PhysRevB.80.094408}. In a perfect crystal, every molecule's easy axis lies along the crystal $c$-axis. However, in Mn$_{12}$-ac, each molecule is surrounded by four acetic acid solvent molecules. Each acetic acid can form only one OH$\ldots$O hydrogen-bond with the two Mn$_{12}$ molecules it lies between. Thus each Mn$_{12}$ molecule can have $n$ ($n=0-4$) hydrogen-bonds around it, which results in six different isomers  \cite{PhysRevLett.91.047203}, and three different easy axis tilts (the molecule’s easy-axis forms an angle with the crystal c-axis). So although the molecule itself has $S_{4}$  symmetry, the Mn$_{12}$-ac does not retain this symmetry in a crystal. As explained in detail later, in the absence of an applied magnetic field these small tilts of the magnetic easy axis of Mn$_{12}$-ac molecules have negligible effect on the magnetic properties. However, when a field is applied these small tilts can play an important role.

\section{Experiment}
{\subsection{Sample Preparation  and Measurement Techniques}
Measurements of the longitudinal magnetization and susceptibility were performed on three Mn$_{12}$-ac-MeOH single crystals of dimensions $\sim 0.2 \times 0.2 \times 0.95$ mm$^3$, $0.085 \times 0.085 \times 0.68$ mm$^3$, and $0.075 \times 0.075 \times 0.85$ mm$^3$ (samples A, B and C, respectively) and three Mn$_{12}$-ac crystals of dimensions $\sim 0.4 \times 0.4 \times 2.17$ mm$^3$, $\sim 0.4 \times 0.4 \times 2.4$ mm$^3$ and $0.3 \times 0.3 \times 1.85$ mm$^3$ (crystals D, E and F, respectively). Details on sample preparation and handling are described on Refs\,\cite{Subedi2010} and \cite{Subedi2012}.  A Hall sensor, (active area $20 \times 100\ \mu$m$^2$) was placed at the end of the crystal and used to measure the magnetization, $M_z$, along the easy direction (\textit{c}-axis) of the crystal (see the bottom inset of Fig. \ref{steps}). All measurements were taken between $0.5$ K and $6$ K in a $^3$He refrigerator in a 3D vector superconducting magnet. A small longitudinal field, $H_z$, was swept along the sample's easy axis at rates between $1 \times 10^{-5}$ T/s and $6.7 \times 10^{-3}$ T/s, in the presence of a series of fixed transverse fields, $H_\perp$ applied in the $y$ direction. For details on measurement setup and techniques see Refs\,\cite{Subedi2010} and \cite{Subedi2012}.

\begin{figure}[t]
\centering
\includegraphics[width=\linewidth]{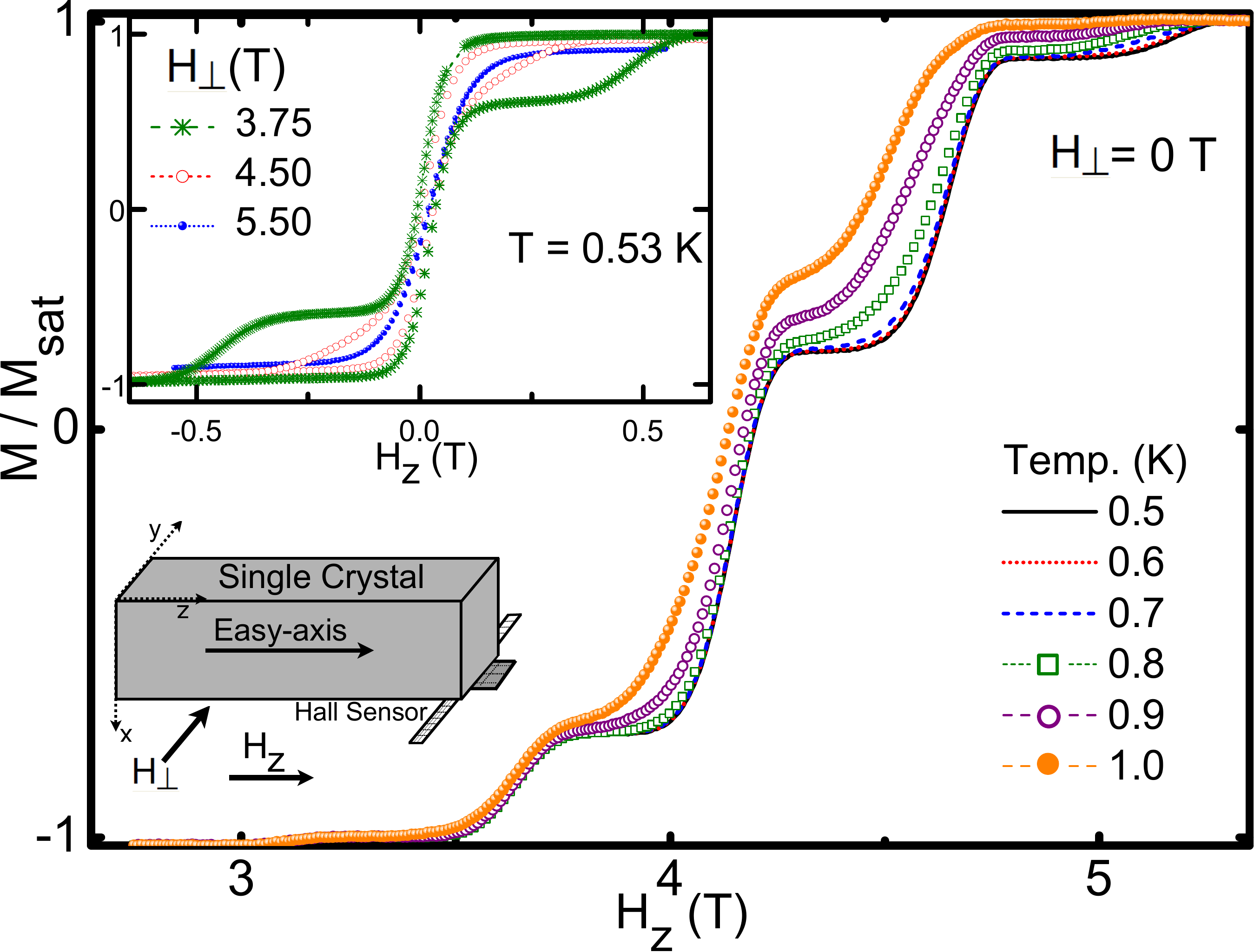}
\caption{Normalized magnetization of Mn$_{12}$-ac-MeOH as a function of longitudinal magnetic field, $H_z$, in zero transverse field at several temperatures below $1$ K. The sweep rate of $H_z$ in the main panel and the top inset is 1.67 mT/s. Top Inset: Magnetization vs  $H_z$ at $T = 0.53$ K for different $H_\perp$.  Bottom Inset: Schematic diagram of the sample, the Hall sensor and magnetic fields. Reprinted from Ref\,\cite{Subedi2012} (\copyright APS).}
\label{steps}
\end{figure}

\subsection{Measurement of the equilibrium susceptibility}

Due to the anisotropy barrier, magnetic hysteresis occurs below a blocking temperature $T_{B}$, of approximately $3$ K in zero field. This precludes measurements of equilibrium properties at temperatures where magnetic ordering is predicted. Therefore, our approach was to measure the magnetic susceptibilities above the blocking temperature and deduce the nature of the magnetic interactions from fits to the Curie-Weiss law. Hence, to measure the susceptibility, we need to make sure that the sample has achieved equilibrium before we can obtain a valid measurement. The factors that determine whether the measurement is taken at equilibrium are the longitudinal field sweep rate, $\alpha$, the temperature $T$ and the transverse field  $H_\perp$. However, $T$ and $H_\perp$ are also parameters affecting the susceptibility; this leaves $\alpha$ as the only control parameter that can be adjusted to reach equilibrium. At low temperature, for a given $H_\perp$, the relaxation time gets longer; hence, we need to keep $\alpha$ low enough so that the experimental time scale stays larger than the relaxation rate. In order to establish the range of experimental conditions in which it is possible to measure the equilibrium susceptibility we first determine the blocking temperature $T_{B}$ as a function of transverse field.

In the main panel of Fig.\ref{steps}, the magnetization of Mn$_{12}$-ac-MeOH (Sample C) in the absence of transverse magnetic field is shown for temperatures below $1$ K. The steps in the magnetization occur due to faster spin-reversal due to resonant tunneling of magnetization when energy-levels on opposite spin-projections coincide at specific magnetic fields \cite{JonathanPRL1996}. In Mn$_{12}$-ac-MeOH the resonant fields at which the steps occur are the same as in Mn$_{12}$-ac, which indicates that the two systems have similar spin energy-level structures. The effect of transverse field is demonstrated in the inset of Fig. 1. The transverse field promotes quantum tunneling by mixing the eigenstates of S$_z$,  and accelerates relaxation toward equilibrium. At 0.53 K, for H$_{\bot}$ = 3.75 T (green stars), the magnetization curve has hysteresis. Increasing the transverse field decreases the area of the loop and at the higher field of H$_\bot$ = 5.5 T (blue solid dots), there is no hysteresis--the sample has achieved equilibrium.  

\begin{figure}[t]
\centering
\includegraphics[width=\linewidth]{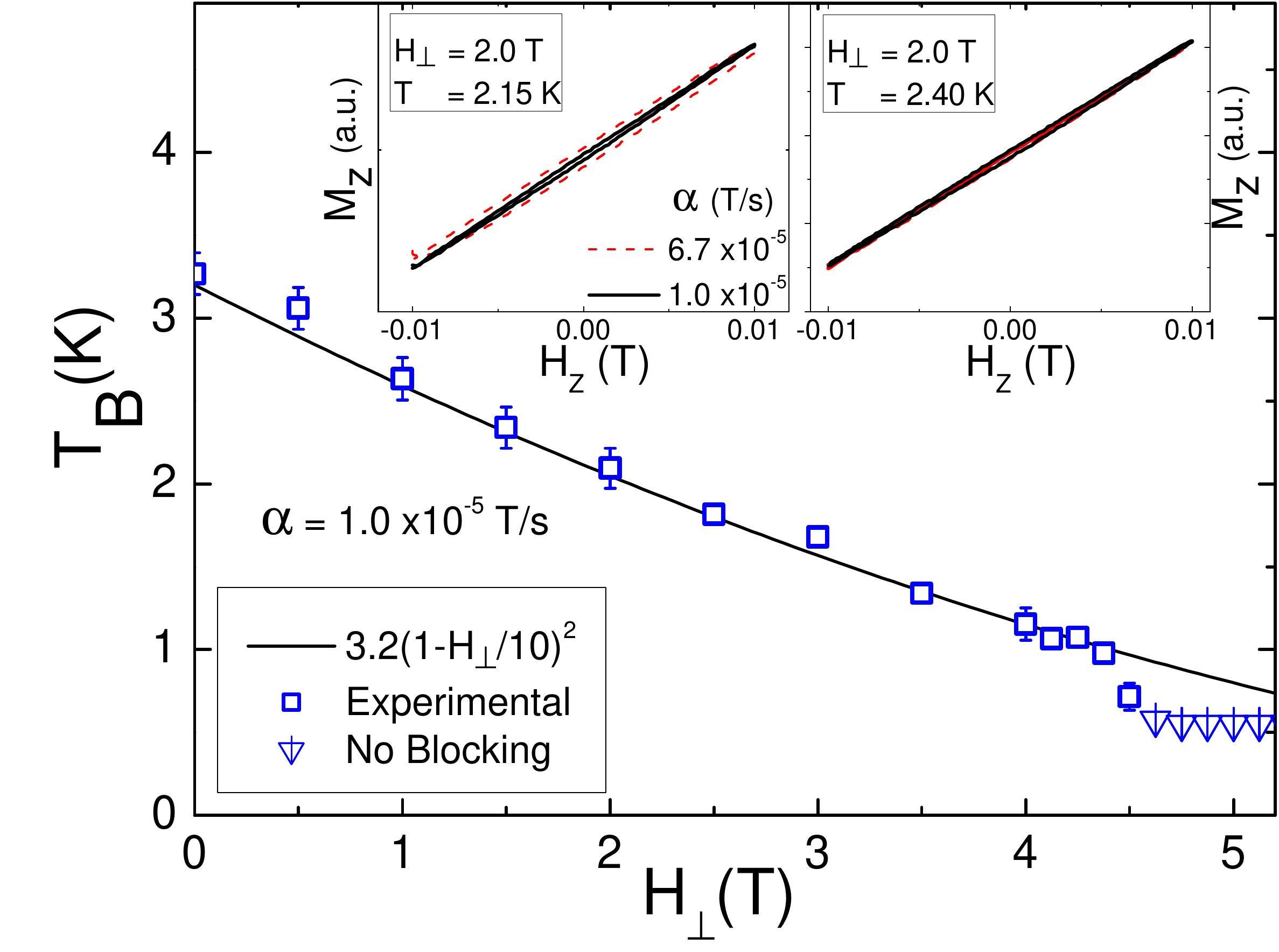}
\caption{Determination of the blocking temperature. Inset: Magnetization as a
function of the longitudinal field swept at the indicated rates for $H_\perp
= 2$ T at $T = 2.15$ K and $T = 2.40$ K. There is a very small amount of temperature independent hysteresis due to the flux trapped in the magnet that is treated as a background. Main panel: Blocking temperature as a function of $H_\perp$. }
\label{tb}
\end{figure}

The hysteresis can also be eliminated by sweeping the longitudinal field sufficiently slowly. The effect of reducing the sweep rate $\alpha$ is demonstrated in the two insets of Fig. \ref{tb} which show the hysteresis loops obtained for two different longitudinal magnetic field sweep rates in a narrow range $\pm 0.01$ T about $H_{z}=0$, measured in the presence of a constant transverse field $H_{\perp}=2$ T at $T=2.15$ K and $T=2.40$ K. At 2.15 K, hysteresis is observed at $\alpha = 6.7 \times 10^{-5}$ T/s indicating that the system is below the blocking temperature; at the slower sweep rate, $\alpha = 1.0 \times 10^{-5}$ T/s  the hysteresis loop is closed, indicating the system is above the blocking temperature. However at 2.40 K, the magnetization is reversible and there is no hysteresis for both sweep rates and equilibrium is reached. From the opening and closing of the hysteresis loop, we deduce the field dependence of $T_{B}$ (Figure \ref{tb}). As expected the applied transverse field $H_{\perp }$ accelerates the relaxation of the magnetization towards equilibrium and thereby lowers the blocking temperature, $T_{B}$. Note that a reduction in $T_{B}$ is also expected from a classical model of single domain uniaxial nanomagnets--the classical version of Mn$_{12}$-ac--where $T_{B}=(1-h)^{2}$, $h=H_\perp/H_{A}$, $H_\perp$ is the externally applied transverse field and $H_{A}$ is the anisotropy field ($H_{A}=2DS/g\mu _{B}\approx 10$ T) \cite{Friedman2}. The solid line in\ Figure \ref{tb} is a fit of the measured $T_{B}$ to the predicted quadratic dependence on field. However, we find that above $H_\perp = 4.5 $ T, at our base temperature of 0.5 K, the system is not blocked and $T_B$ deviates from the classical expectations. This is associated with the onset of pure (i.e. temperature independent) quantum tunneling \cite{Bokacheva2000}.

\subsection{Longitudinal susceptibility}

\begin{figure}[thb]
\centering
\includegraphics[width=\linewidth]{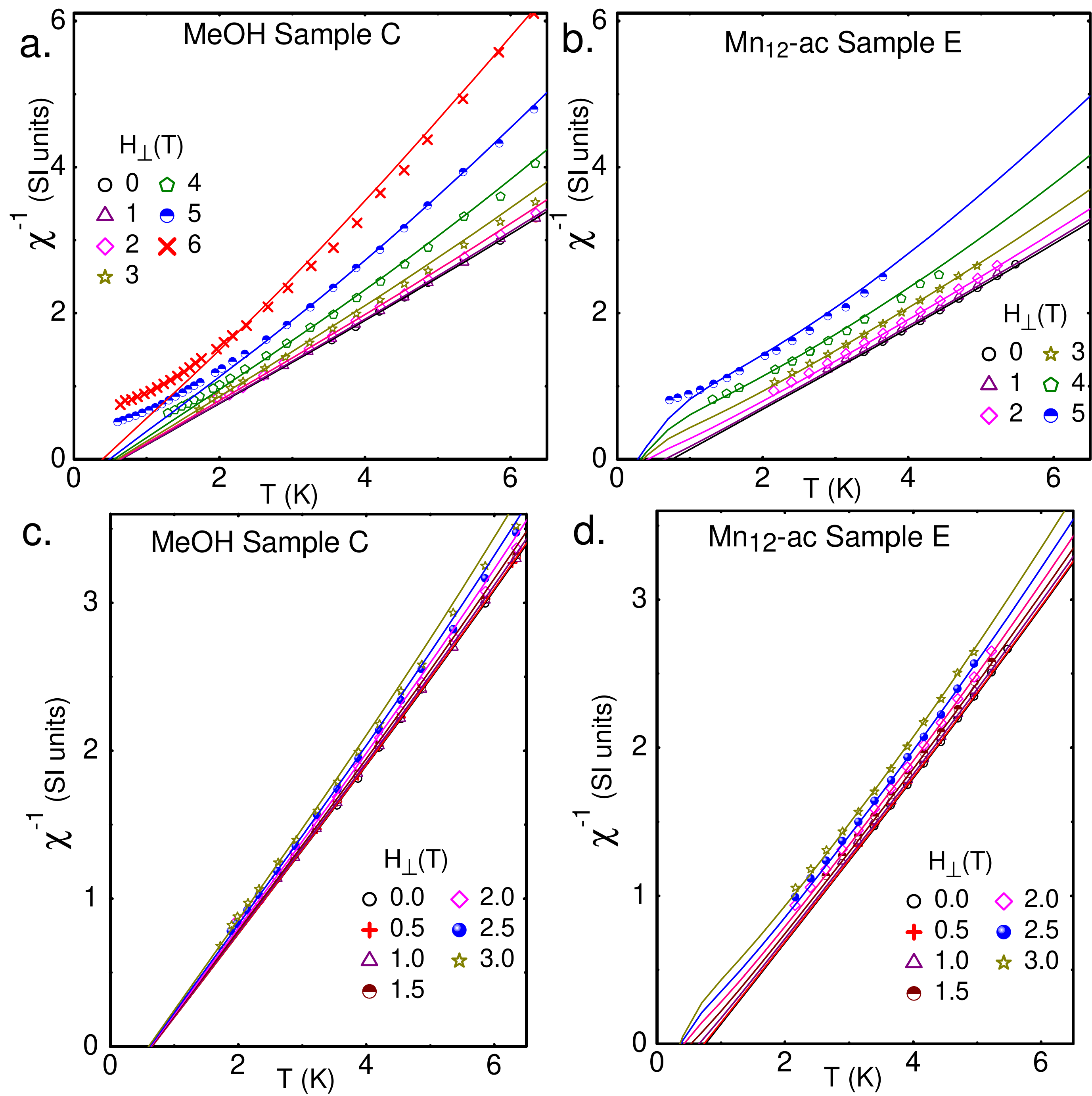}\caption{(a): Inverse susceptibility as a function of temperature for Mn$_{12}$-ac-MeOH (Sample C) in various transverse fields up to $6$ T. (b): Inverse susceptibility as a function of temperature of Mn$_{12}$-ac (Sample E) for various transverse fields up to $5$ T.  (c) and (d) Inverse susceptibility in low transverse field up to 3 T for  Mn$_{12}$-ac-MeOH and Mn$_{12}$-ac, respectively.  The solid lines are theoretical curves based on Eq.\,\ref{Millis32}. Reprinted from Ref\,\cite{Subedi2012} (\copyright APS).}
\label{fig2}
\end{figure}

Figure \ref{fig2} presents the measured equilibrium longitudinal susceptibility, $\chi\equiv\partial M_z/\partial H_z|_{H_z=0}$, of Mn$_{12}$-ac-MeOH and Mn$_{12}$-ac single crystal plotted as $\chi^{-1}$ versus temperature. Figures \ref{fig2}(a) and \ref{fig2}(c) show the inverse susceptibility of a Mn$_{12}$-ac-MeOH (Sample C) crystal as a function of temperature for various fixed transverse magnetic fields between 0 and 6 T. For zero transverse field $\chi^{-1}$ obeys the Curie-Weiss law, $\chi ^{-1}\sim (T-T_{CW})$, expected from mean field theory (MFT).  The  intercept $T_{CW}$ implies a transition at this temperature from paramagnetism (PM) to ferromagnetism (FM). As we apply the transverse field, there is a systematic increase in the inverse susceptibility, accompanied by a progressively larger deviation from the straight-line behavior found at $H_\perp=0$ due to canting of spins in the transverse field. 

Similar data was obtained for a Mn$_{12}$-ac (Sample E) crystal and is shown in Figs. 3(b) and 3(d). In Mn$_{12}$-ac, the zero field intercept  is found to be $T_{CW}\sim 0.9$ K consistent with the result of Luis \textit{et al.} \cite{LuisPRL2005}.  The expected overall decrease of the susceptibility with transverse field is also observed, as well as the anomalous deviations at low temperature.  However, as demonstrated by the data at small transverse fields in fig \ref{fig2} (c) and (d), the response of $\chi$ to transverse field is distinctly different for the two systems. The slopes of the $\chi^{-1}$ vs. $T$ curves increase rapidly with the transverse field for Mn$_{12}$-ac-MeOH; however,  in Mn$_{12}$-ac, the curves remain approximately parallel with little change of slope  and the apparent Weiss temperature decrease rapidly with the increasing transverse field \cite{Subedi2010}.

\section{Theory}
\subsection{TFIFM}
The application of a magnetic field in a direction transverse to the Ising axis induces quantum spin fluctuations that compete with the long-range order by mixing the eigenstates of $S_z$ \cite{FriedmanJAP1997}. A fundamental model used to study the interplay between the long range order and spin fluctuations is the Transverse-Field Ising Ferromagnet Model (TFIFM), which is described by following Hamiltonian:
\begin{equation}
 \mathcal{H}= \frac{1}{2} \sum_{i\neq j} J_{ij} S^z_i S^z_j+  \Delta \sum_iS_i^x
 \label{TFI}
 \end{equation}
where, $S_i$ is a two level Ising spin on lattice site $i$, $J_{ij}$ are the dipolar couplings and $\Delta$ is the tunnel splitting that depends on the applied transverse field \cite{Millis09}. Since Mn$_{12}$-ac-MeOH and Mn$_{12}$-ac are chemically similar, we expect to observe the same magnetic response with the transverse field in both systems. However, we find that the behavior of Mn$_{12}$-ac-MeOH is consistent with the Transverse Field Ising Ferromagnet Model, Eq.~\ref{TFI}. Whereas, we observe more rapid suppression of ferromagnetism by the transverse field in Mn$_{12}$-acetate than is predicted by this same model (Fig. 3b, 3d). As described in Section 2.2, the two systems differ only in the solvent molecules and that apparently give rise to different response. 

\subsection{Randomness in Mn$_{12}$-ac}
The large change in the susceptibility in relatively small transverse fields and the rapid suppression of ferromagnetism in Mn$_{12}$-ac (Fig. 3b, 3d) indicates that the TFIFM fails to describe the data. This suggests the presence of physics not included in this model.  We argue that the additional physics is a random-field effect arising from structural disorder in the Mn$_{12}$-ac crystal.  As described earlier, different isomers of the host acetate material have been shown \cite{Park02} to cause the spin quantization axis of some of the Mn$_{12}$-ac molecules to tilt away from the crystal $z$-axis by a small  monomer-dependent angle ($\theta \approx \pm 1^\circ$) \cite{PhysRevLett.91.047203,PhysRevLett.90.217204,PhysRevB.70.094429,AndyJLTP2005,Park04}. As illustrated schematically in Fig. \ref{spin}, in the absence of an applied magnetic field these small tilts of the magnetic easy axis of individual molecules have negligible effect on the magnetic properties. However, a field applied transverse to the easy (Ising) axis has a nonvanishing projection along the tilted local spin quantization axis. This additional field becomes comparable in magnitude to the dipolar field itself for transverse field magnitudes of order $4$ T \cite{Millis09, Subedi2010}. The distribution of the isomers in the crystal is assumed to be random and thus, the transverse field introduces random longitudinal fields that further supress the long range order. 

\begin{figure}[h]
\centering
\includegraphics[width=\linewidth]{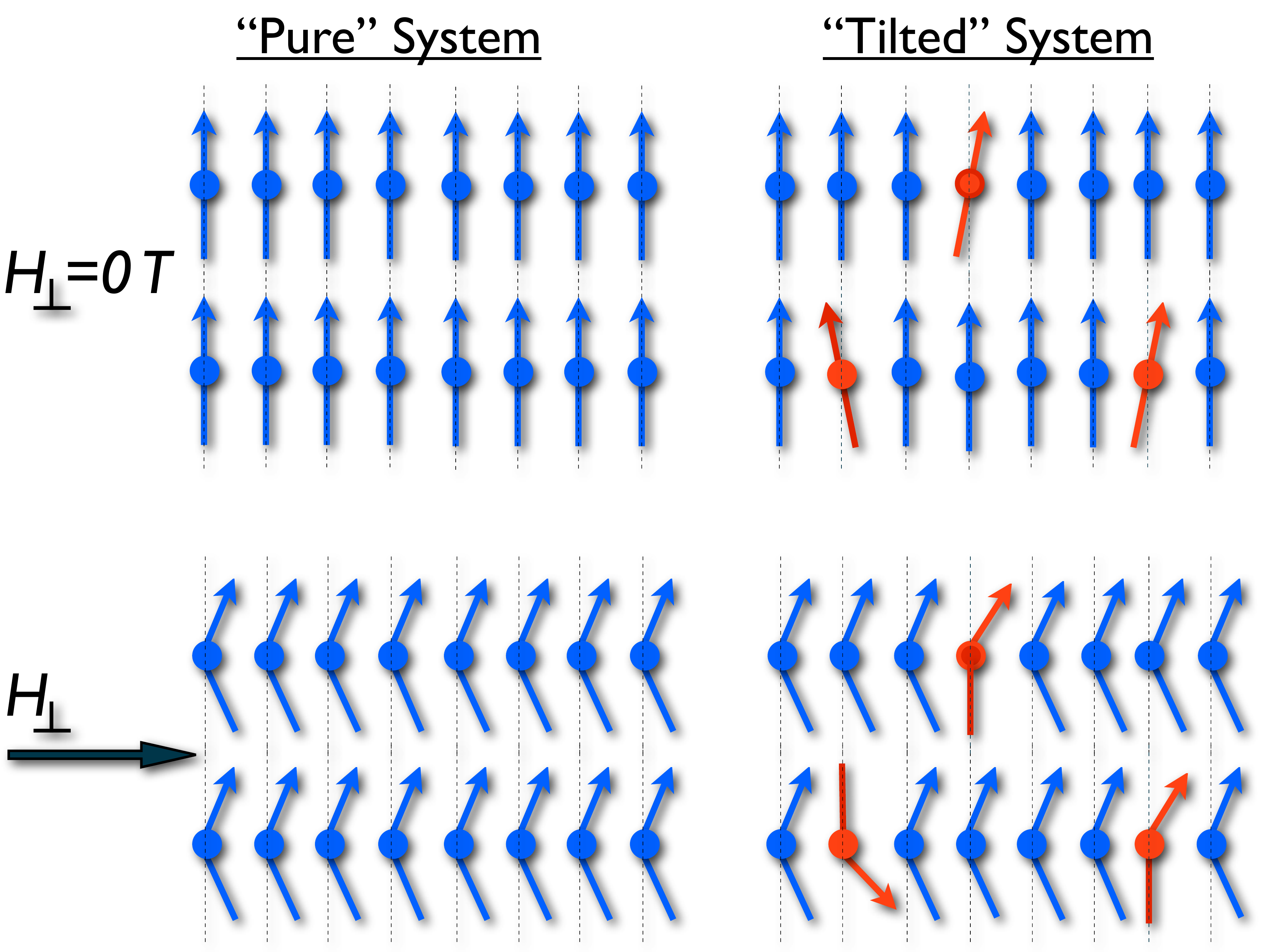}
\caption{Effect of easy axis tilts on the transition temperature. In zero field a perfectly ordered crystal and a crystal in which there are easy axis tilts (e.g., the red spins) will order at nearly the same temperature (the small tilts do not greatly modify the interaction between spins, which depends on the longitudinal component of the magnetic moment). In an applied transverse field, the spins of misaligned molecules experience a field along their Ising axis. When this field is comparable to the exchange field these spins are frozen (red spins) and do not order. This leads to an effective dilution of the spins, a decrease in the susceptibility and a reduction in the transition temperature. It also increases the random-field on the other sites in the crystal.}
\label{spin}
\end{figure}

\subsection{RFIFM}
We can account for the random-field by adding a transverse field-dependent random-field term $h_{i}$ to Eq.~\ref{TFI} and get the Hamiltonian that describes the Random-Field Ising Ferromagnet (RFIFM): 
\begin{equation}
 \mathcal{H}= \frac{1}{2} \sum_{i\neq j} J_{ij} S^z_i S^z_j+  \Delta \sum_iS_i^x+\sum_{i} h_{i} S^z_i
 \label{RFIFM}
 \end{equation}
where, in the small polar tilt limit, $h_{i} = \theta _{i}cos(\phi _{i}+\phi _{H})g\mu _{B}H_{\bot}$ is the site-dependent field that arises from the  misorientation of the Mn$_{12}$-ac spin-quantization (spin-z) axis with respect to the crystallographic $z$ axis \cite{Park04}. The misorientation is characterized by a small polar tilt $\theta $ and an azimuthal angle $\phi $, which is measured with respect to the direction $\phi _{H}$ defined by the transverse field. Setting $\theta$ and $\phi$ to be zero would result in vanishing $h_i$ term and the resulting model is TFIFM.

\subsection{Susceptibility}
We use mean field theory (MFT) to calculate the inverse magnetic susceptibility using the above Hamiltonian (Eq\,\ref{RFIFM}, for detail calculations see Ref.~\cite{Millis09}) and get:
\begin{equation}
\chi^{-1}(T,H_T,z)=C\left(\frac{T^2+\frac{1}{3} \Delta^2(H_\bot)+\left<h_i^2\right>}{Tcos^2\theta_c} \right)-J .
\label{Millis32}
\end{equation}
Here $J$ is the effective exchange interaction obtained from the appropriate spatial average over the dipole interaction, the angle $\theta_c$ characterizes the spin canting in an applied transverse field, $C$ represents sample volume,  and $\Delta(H_\bot)$ is the tunnel splitting; detailed expressions for the  dependence of $\theta_c$ and $\Delta$ on $H_\bot$ are given in Ref.~\cite{Millis09}. $\left<h_i^2\right>$ is the mean square amplitude of the random-field distribution. Calculations were performed using the monomer distribution proposed by Park \cite{Park04} which involves twelve inequivalent sites randomly distributed through the crystal. For temperatures greater than $2$ K, the only important feature is the mean square amplitude of the tilt angle and changing the form of the distribution but keeping the amplitude of the tilt fixed leaves the results invariant.  For lower temperatures, specific properties of the distribution become important, in particular the fraction of sites with vanishing random field. In the Park model the angles $\phi_i=n\pi/4$ with $n=0,1,2...7$ and the fraction of sites with very small random field may be controlled by varying $\phi_H$.

\section{Discussion}
The solid lines shown in Fig. \ref{fig2} (a) and (c) are the result of calculations for TFIFM (setting $\theta$ and $\phi$ to be zero, i.e. eliminating $\left<h_i^2\right>$ in Eq.~\ref{Millis32}).  As $H_{\perp }$ is increased, the slope of the calculated traces increases, reflecting spin canting induced by the magnetic field. Also, due to the increase in quantum tunneling, the estimated ferromagnetic transition temperature (the extrapolated value where $\chi ^{-1}$ vanishes) decreases and the calculated traces develop a weak curvature at low temperatures. The TFIFM calculation and the Mn$_{12}$-ac-MeOH data agree well. 

The solid lines shown in Fig. \ref{fig2} (b) and (d) show the results of calculations that include the randomness associated with isomer tilts.  Here we have assumed that all the sites have a tilt and the mean square tilt angle is $1.8^{\circ }$, larger than the $0.4^{\circ }$ calculated by Park \textit{et al.} \cite{Park02} but in good accord with the values determined using EPR \cite{PhysRevB.70.094429} (which finds tilts up to 1.8$^{\circ }$). The theory with randomness accounts for all the major features of the observed $\chi ^{-1}(T)$ in Mn$_{12}$-ac in the transverse fields.

The $\chi^{-1}(T)$ data between 2 and 6 K in Fig. \ref{fig2} are consistent with theory for both samples, where a particularly good fit is obtained for fields below 3 T. However the plots in Fig. \ref{fig2} reveal a discrepancy at low temperatures and high fields. While the theoretically calculated lines intersect the temperature axis at $T_{CW}(H_\bot)$ implying the approach to a ferromagnetic phase, the measured susceptibility deviates from this simple behavior, flattening as the temperature decreases toward the presumed transition. The behavior observed at these low temperatures and high fields is not understood, and may imply that a transition to a new phase is prohibited for reasons that are unclear. 

For both systems with and without randomness, a transverse field leads to a canting of the spins away from the $z$-axis, as illustrated in Fig. \ref{spin}, and enhances quantum fluctuations of the spin. The strength of intermolecular dipole interaction, which is associated with the z-component of spin, gets reduced due to canting. A transverse field much smaller than the anisotropy field $H_A\approx 10$ T produces very little spin canting ($\tan\theta_c =H_\perp/H_A$) and thereby a negligible change in the interaction strength. As a result, the susceptibility and the ordering temperature are virtually unchanged in the system without randomness for transverse fields below 3T. Only when the tunnel splitting, $\Delta$, of the lowest spin states is comparable to the intermolecular dipole interactions, the quantum fluctuations become important and the FM order is very strongly suppressed (when the transverse field $H_\perp$ approaches the anisotropy field $H_A$, the energy eigenstates are going from the eigenstates of $S_z$ to eigenstates in the direction of the field ($S_x$)). This occurs at $\sim7$ T for both Mn$_{12}$-ac-MeOH and Mn$_{12}$-ac \cite{garanin:174425,Millis09}.  However, due to the tilts present in the disordered system,  transverse field gives rise to the random-fields comparable in magnitude to the intermolecular dipole field (approximately 50 mT \cite{McHugh2009}, corresponding to 3 T for a tilt angle of 1$^{\circ}$). In such system, the tilted spins can no longer participate in the FM order and there is an effective dilution of the spins which causes a rapid reduction of the susceptibility and of the ordering temperature as seen in the disordered system of Mn$_{12}$-ac.
\cite{Millis09}.

\begin{figure}[h]
\centering
\includegraphics[width=\linewidth]{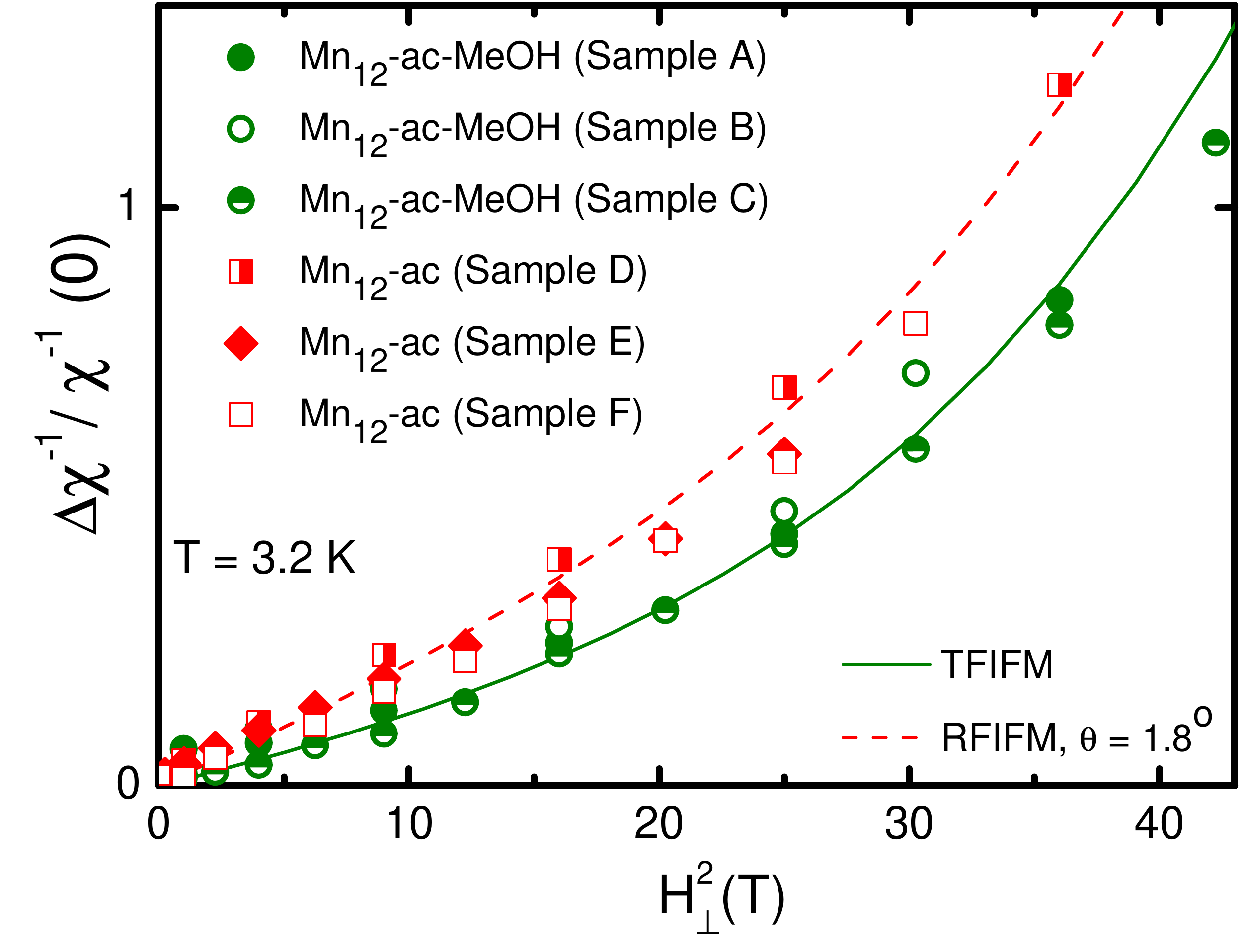}
\caption {(Color on line) The change in inverse susceptibility, $\Delta\chi^{-1}$, normalized to the susceptibility at zero field versus $H_\perp^2$ for Mn$_{12}$-ac-MeOH (green dots) and Mn$_{12}$-ac (red squares) at $T = 3.2$ K. The red dashed line is calculated using RFIFM for the root mean square tilt angle of $1.8\,^{\circ}$. The solid green line shows the result for TFIFM. Reprinted from Ref\,\cite{Subedi2012} (\copyright APS).}
\label{fig3}
\end{figure}

Guided by  Eq.~\ref{Millis32},  we further demonstrate  the different response to the magnetic field in Fig.\,\ref{fig3} by plotting the normalized change of the inverse susceptibility,
\begin{equation}
\Delta\chi^{-1} (H_\perp)=\frac{\chi^{-1}(H_\perp)-\chi^{-1}(0)}{\chi^{-1} (0)},
\end{equation} 
as a function of $H_\perp^2$ for three Mn$_{12}$-ac-MeOH samples (green dots) and three Mn$_{12}$-ac (red squares) at T = 3.2 K.  We note that the subtraction, $\Delta \chi^{-1}(H_\perp) = \chi^{-1}(H_\perp) - \chi^{-1} (0)$, eliminates the intermolecular interaction term ($J$), and the normalization removes the dependence on sample volume, $C$. Therefore, $\Delta\chi^{-1}$ only shows the effect of transverse field. Figure  ~\ref{fig3} clearly shows that the effect of the transverse field is much larger for Mn$_{12}$-ac consistent with the discussion above regarding the additional effect of transverse field in the system. In the same figure we plot the normalized change of the inverse susceptibility calculated for the TFIFM with no tilt angle (green solid line) and for the RFIFM with the mean square tilt angle of $1.8\,^{\circ}$ (red dashed line) for the random-field distribution proposed by Park et. al ~\cite{Park04}.  The excellent agreement between calculation and data for the MeOH material at $H_\bot<4T$ at $3.2$ K  is an indication that this system is a realization of the dipolar Ising model in a transverse field.  A good fit obtained for the Mn$_{12}$-ac crystal data with the calculations using root mean square tilt angles of $1.8\,^{\circ}$ suggest that the material is a realization of  dipolar random-feld Ising model in a transverse field.

\begin{figure}[h]
\includegraphics[width=\linewidth]{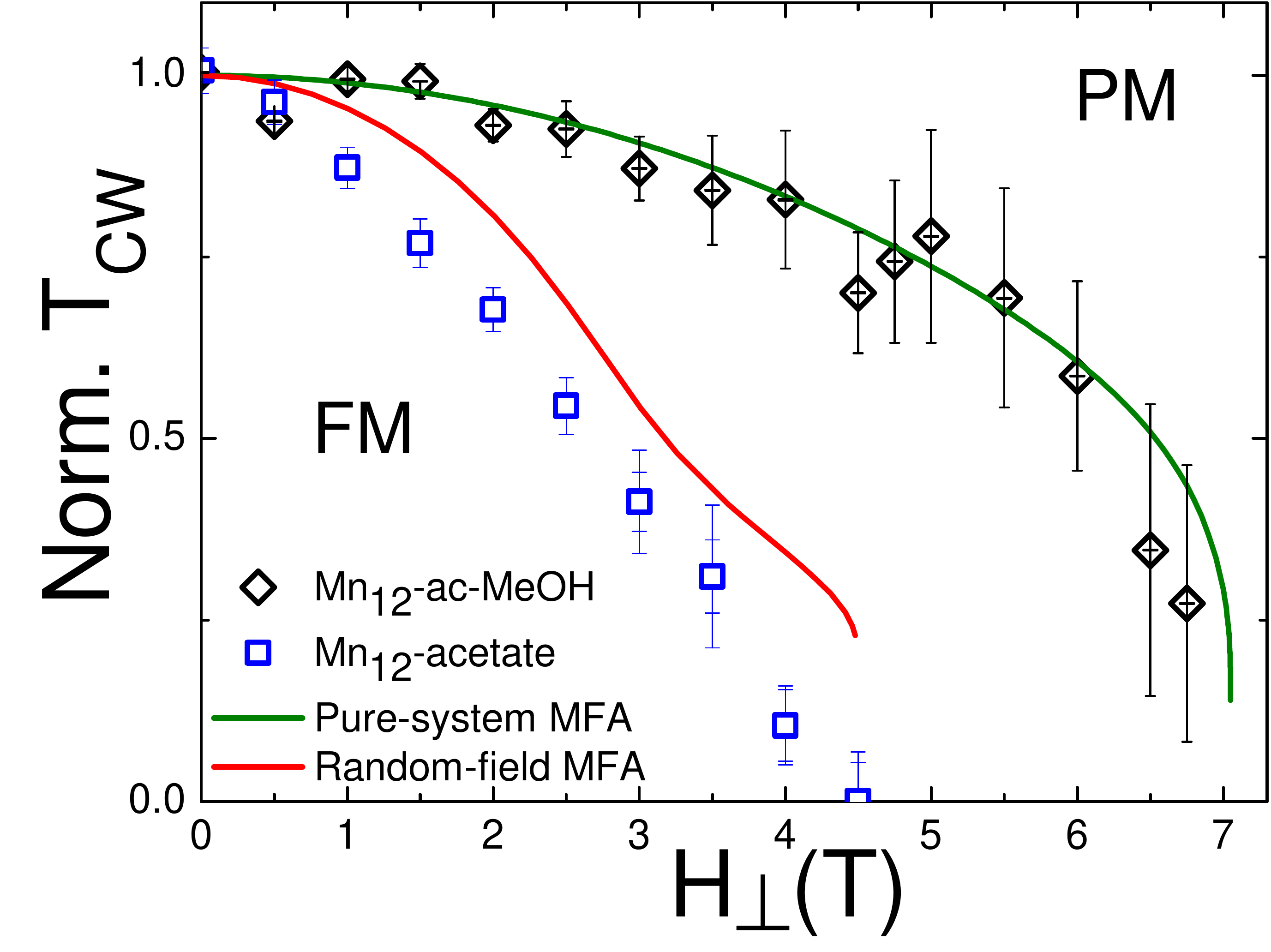}
\caption{The Curie-Weiss and the ferromagnetic transition temperatures as a
function of transverse field. The intercepts $T_{CW}$ of  Mn$_{12}$-ac-MeOH (black diamonds)  and Mn$_{12}$-ac (blue squares) are
obtained from the straight-line portion of the data curves in Figure \ref{fig2} (a) and \ref{fig2} (b). The green and red solid lines are mean-field transition temperatures, $T_{C}$, calculated for the pure and
random case, respectively.  $T_{C}$ for RFIFM is calculated using the distribution obtained by Park \protect\cite{Park04} and the parameters ($\protect\theta =1.8^{\circ }$ and $\protect\phi =0$) are the same as those used to fit the data in Figure \protect\ref{fig2}.}
\label{phase}\centering
\end{figure}

Figure \ref{phase} shows approximate values of the the Curie-Weiss intercepts $T_{CW}$ of  Mn$_{12}$-ac-MeOH (black diamonds)  and Mn$_{12}$-ac (blue squares) obtained from fitting the high-temperature region of the experimental curves shown in Figure \ref{fig2} (a) and \ref{fig2} (b) to the Curie-Weiss law. The fit does not include data at low temperature and higher fields where, as discussed above, $\chi^{-1}$ deviates from the Curie-Weiss behavior. The initial  suppression of $T_{CW}$ in Mn$_{12}$-ac-MeOH for transverse fields $H_\bot<5$ T is expected due to spin canting, which reduces the net moment in the axial direction; the more rapid suppression at higher fields derives from the tunnellng term, $\Delta$. A substantially more rapid suppression of $T_{CW}$ with $H_\bot$ is evident for Mn$_{12}$-ac. The results for Mn$_{12}$-ac are consistent with a modified theory that includes the effects of random fields arising from the tilt angles.  We also plot the calculated the mean-field PM-FM transition temperature $T_{C}$ (where $\chi$ diverges). The calculated values of $T_{C}$\ for the TFIFM and RFIFM are denoted in Fig. \ref{phase} by the solid green and red lines, respectively. For the theoretical calculation, the dipolar part of the interaction was obtained using the measured lattice parameters and crystal structure of  Mn$_{12}$-ac (very similar values of Mn$_{12}$-ac-MeOH) and the spin canting and tunnel splitting were obtained as described in Ref.~\cite{Millis09}. 

\section{Conclusion}
In summary, these studies demonstrate that the temperature dependence of the susceptibility and the dependence of the (extrapolated) Weiss temperature on applied transverse field are different for the two very closely related materials.  We observe that the magnetic susceptibility of Mn$_{12}$-ac-MeOH follows the behavior expected for a transverse field Ising ferromagnet (TFIFM), in marked contrast with Mn$_{12}$-ac, which is a new archetype of random-field Ising ferromagnetism in transverse field. In this system, although the intrinsic randomness in the interaction is small, it is sufficient for an externally applied transverse magnetic field to generate a significant random field in the longitudinal direction. More broadly, the availability of these two very chemically similar SMMs with distinct types of magnetism provides unique opportunities for experimental studies of the effect of randomness on quantum phase transitions and magnetic relaxation.

\section{Acknowledgement}
\label{acknow}

It is a pleasure to contribute to this special issue of Polyhedron in honor of Prof. George Christou, with whom we have had a long-standing and very productive collaboration. We have published many joint papers over the years, including the research on Mn$_{12}$-ac that we review in this article [see References \cite{Subedi2010} and \cite{Subedi2012}]. ADK and PS acknowledge support by ARO W911NF-08-1-0364 and NYU-Poly Seed Funds; MPS acknowledges support from NSF- DMR-0451605; YY acknowledges support of the Deutsche Forschungsgemeinschaft through a DIP project. AJM acknowledges support of NSF-DMR-1006282.


\end{document}